 \documentstyle[11pt,aaspp4]{article}
 \slugcomment{Accepted for publication in
                              {\it The Astrophysical Journal\/} 04/17/98}
\lefthead{Mauche \& Raymond}
\righthead{ORFEUS Spectra of AM Her}

\def\Mwd  {M_{\rm wd}}
\def\Rwd  {R_{\rm wd}}

\def\Msun {{\rm M_{\odot}}}
\newcommand\lax{{\lower0.75ex\hbox{$<$}\atop\raise0.5ex\hbox{$\sim$}}}
\newcommand\gax{{\lower0.75ex\hbox{$>$}\atop\raise0.5ex\hbox{$\sim$}}}

\begin{document}

\title{ORFEUS II Far-UV Spectroscopy of AM Herculis}

\author{Christopher W.\ Mauche\altaffilmark{1}}
\affil{Lawrence Livermore National Laboratory, \\
       L-41, P.O.\ Box 808, Livermore, CA 94550; \\
       mauche@cygnus.llnl.gov}

\and

\author{John C.\ Raymond\altaffilmark{1}}
\affil{Harvard-Smithsonian Center for Astrophysics, \\
       60 Garden Street, Cambridge, MA 02138; \\
       jraymond@cfa.harvard.edu}

\altaffiltext{1}{ORFEUS-SPAS II Guest Investigator}

\clearpage % force page break; deleted when you change from pp to ms

% Abstract
%---------------------------------------------------------

\begin{abstract}
Six high-resolution ($\lambda/\Delta\lambda\approx 3000$) far-UV
($\lambda\lambda=910$--1210~\AA ) spectra of the magnetic cataclysmic
variable AM~Herculis were acquired in 1996 November during the flight of
the ORFEUS-SPAS II mission. AM~Her was in a high optical state at the
time of the observations, and the spectra reveal emission lines of O~VI
$\lambda\lambda 1032$, 1038, C~III $\lambda 977$, $\lambda 1176$, and
He~II $\lambda 1085$ superposed on a nearly flat continuum. Continuum
flux variations can be described as per G\"ansicke et~al.\ by a $\approx
20$ kK white dwarf with a $\approx 37$ kK hot spot covering a fraction
$f\sim 0.15$ of the surface of the white dwarf, but we caution that the
expected Lyman absorption lines are not detected. The O~VI emission lines
have broad and narrow component structure similar to that of the optical
emission lines, the C~III and He~II emission lines are dominated by the
broad component, and the radial velocities of these lines are consistent
with an origin of the narrow and broad components in the irradiated face
of the secondary and the accretion funnel, respectively. The density of
the narrow- and broad-line regions is $n_{\rm nlr}\sim 3\times 
10^{10}~\rm cm^{-3}$ and $n_{\rm blr}\sim 1\times 10^{12}~\rm cm^{-3}$,
respectively, yet the narrow-line region is optically thick in the O~VI
line and the broad-line region is optically thin; apparently, the
velocity shear in the broad-line region allows the O~VI photons to
escape, rendering the gas effectively optically thin. Unexplained are
the orbital phase variations of the emission-line fluxes.
\end{abstract}


\clearpage % force page break; delete when you change from pp to ms
 
% Body of the paper
%---------------------------------------------------------

\section{Introduction}

AM~Herculis is the prototype of a class of strongly magnetic ($B\approx
10$--100 MG) cataclysmic variables (see \cite{cro90} for a review and 
the volume by \cite{buc95} for recent results). The strong field locks
the white dwarf into corotation and the accreting matter is channeled
along the field lines for most of its trajectory from the secondary's
inner Lagrange point to the white dwarf magnetic pole. A strong shock
forms in the stream of freely falling matter close to the white dwarf
surface, thermalizing a significant fraction of its kinetic energy and
raising its temperature to a few tens of keV. The accretion region is
therefore a source of hard X-rays, but soft X-rays are also produced by
reprocessing of the hard X-rays in the white dwarf photosphere and by
(with a tip `o the hat to Dr.\ Bengue)
``deep heating'' of the white dwarf photosphere by blobs of material
which penetrate to large optical depths before thermalizing their kinetic
energy (\cite{kui82}). The dominant source of the UV continuum is the 
the white dwarf photosphere, while the optical continuum is produced by
free-free and bound-free emission from the funnel above the accretion
shock and possibly the irradiated face of the secondary (\cite{sch91}).
Sources of IR through FUV emission lines include the irradiated face
of the secondary, the accretion stream, and the accretion funnel. The
contribution of each of these emission regions to the net spectrum can
be distinguished by the phasing and radial velocity amplitudes of the
emission lines, and the phasing and amplitude of continuum flux
variations. Because of the small size and high X-ray luminosities of
these binaries, photoionization of all of these emission regions is
likely to be important, with ionization parameters
$\xi\equiv L/nr^2 =100\, L_{33}/n_{11}r_{10}^2$, where $L_{33}$ is the
X-ray luminosity in units of $10^{33}~\rm ergs~s^{-1}$,
$n_{11}$ is the particle density in units of $10^{11}~\rm cm^{-3}$, and
$r_{10}$ is the distance from the X-ray emission region in units of
$10^{10}$~cm.

AM~Her has been observed in the UV numerous times with {\it IUE\/}
(\cite{ray79}; \cite{hei88}; \cite{gan95}; \cite{sil96}; \cite{gan98})
but on only two previous occasions in the FUV: once in 1993 September
during the flight of the ORFEUS-SPAS I mission (\cite{ray95}) and once in
1995 March during the Astro-2 mission (\cite{gre96}). We here discuss six
FUV spectra of AM~Her acquired in 1996 November during the ORFEUS-SPAS II
mission. These observations are superior to those of ORFEUS~I because of
the extensive phase coverage and, thanks to the $\approx 100$\%
improvement in the effective area (\cite{hur96}; \cite{hur97}), the 
higher signal-to-noise ratio of the individual spectra. Combined with the
results from recent EUV and X-ray observations of AM~Her, the ORFEUS~II
spectra allow a detailed picture to be drawn of the locations of, and
physical conditions in, the various FUV emission regions of this compact
magnetic binary.

\section{Observations}

The FUV spectra were acquired with the Berkeley spectrograph in the
ORFEUS telescope during the flight of the ORFEUS-SPAS II mission in 1996 
November--December. The general design of the spectrograph is discussed
by Hurwitz \& Bowyer (1986, 1996), while calibration and performance of
the ORFEUS-SPAS II mission are described by Hurwitz et~al.\ (1997); for
the present purposes, it sufficient to note that spectra cover the range
$\lambda\lambda = 910$--1210~\AA \ and that the mean instrument profile
$\rm FWHM\approx 0.33$~\AA , hence $\lambda/\Delta\lambda\approx 3000$.
Because the orbital period of the ORFEUS-SPAS satellite (91 min) was very
nearly half of the binary orbital period of AM~Her (186 min), to obtain
full phase coverage it was necessary to observe the source in pairs of
consecutive orbits over an interval of 2 days or 17 binary orbits.
Specifics of the observations are collected in Table~1: the HJD of the
start of the exposures, the length of the exposures, and the range of
binary phases assuming Tapia's linear polarization ephemeris
(\cite{hei88}). AAVSO measurements near the time of these exposures
confirm that AM~Her was in a high state, at an optical magnitude of
$13.1\pm 0.2$ (\cite{mat97}).

\section{Results and Analysis}

Figure~1 shows the background-subtracted and flux-calibrated spectra
binned to a resolution of 0.1~\AA \ and smoothed with a 5-point triangular
filter. Relatively strong residual geocoronal emission lines of H~I
$\lambda 1025.7$ (Lyman $\beta $), He~I $\lambda 584.3$ (at 1168.7~\AA \
in second order), N~I $\lambda 1134$, $\lambda 1200$, and O~I $\lambda
988.7$ have been subtracted from these spectra by fitting Gaussians in
the neighborhood ($\pm 5$~\AA ) of each line. The remaining geocoronal
emission lines are all very weak and contaminate only a limited number of
discrete ($\rm FWHM\approx 0.8$~\AA ) portions of the spectra. These FUV
spectra are generally consistent with the two spectra acquired during
the ORFEUS~I mission in 1993 September (\cite{ray95}), with emission
lines of O~VI $\lambda\lambda 1032$, 1038, C~III $\lambda 977$, $\lambda
1176$, and He~II $\lambda 1085$ superposed on a nearly flat continuum.
The broad and variable emission feature at $\lambda\approx 990$~\AA \ may
be N~III, but the flux and position of this feature are uncertain because
it coincides with a strong increase in the background at $\lambda\approx
1000$~\AA \ which renders noisy the short-wavelength end of these spectra.

\subsection{Continuum}

The extensive phase coverage of these observations allows studies of
the phase variability of the FUV continuum and emission lines which were
not possible with the limited amount of ORFEUS~I data. To quantify the
continuum flux density variations, we measured the mean flux density at
$\lambda = 1010\pm 5$~\AA . This choice for the continuum bandpass is
somewhat arbitrary, but it avoids the noisy portion of the spectra
shortward of $\lambda \approx 1000$~\AA , and it is a local minimum in
the spectra. Ordered by orbital relative phase, the mean flux density
in this bandpass is $f_{\lambda} ({\rm 1010~\AA })=0.126$, 0.181, 0.309,
0.403, 0.423, and $0.233\times 10^{-12}~\rm erg~cm^{-2}~s^{-1}~\AA
^{-1}$. As shown in Figure~2, these flux density variations are
reasonably well fitted ($\rm \chi ^2/dof = 8.9/3$ assuming 5\% errors
in the flux densities) by
$f_{\lambda} ({\rm 1010~\AA }) = A + B\,\sin 2\pi (\phi- \phi_0)$ with
$A=0.274\, (\pm 0.006) \times 10^{-12}~\rm erg~cm^{-2}~s^{-1}~\AA ^{-1}$,
$B=0.148\, (\pm 0.007) \times 10^{-12}~\rm erg~cm^{-2}~s^{-1}~\AA ^{-1}$,
and $\phi_0=0.368\pm 0.008$. Since the ORFEUS bandpass is too narrow to
meaningfully constrain the effective temperature, it is not possible to
uniquely determine the cause of these continuum flux density variations:
they could be due to variations in the effective temperature, variations
in the effective size of the emission region, or some combination of
these. Assuming $\Mwd = 0.75\,\Msun $ ($\Rwd = 7.4\times 10^8$~cm),
$d=75$~pc, and that the entire white dwarf surface radiates with a
blackbody spectrum, the effective temperature varies with phase according
to $T_{\rm eff}({\rm kK})=28.8 +3.4\,\sin 2\pi (\phi-0.368)$. G\"ansicke
et~al.\ (1995) modeled {\it IUE\/} spectra of AM~Her in a high state
with white dwarf model atmosphere spectra and found that the observed
variations of the flux {\it and\/} shape of the UV spectra could be
described by a white dwarf of temperature $T_{\rm wd}\approx 20$~kK and
a hot spot of temperature $T_{\rm spot}\approx 37$~kK with a relative
area $f\equiv A_{\rm spot}/4\pi\Rwd ^2\sim 0.08$. If we similarly assume
that we are seeing a 20~kK white dwarf with 37~kK spot, the apparent
projected area of the spot varies with binary phase according to
$f=0.090+0.059\,\sin 2\pi(\phi - 0.368)$. To demonstrate that such
two-temperature blackbody models do a good job of matching the ORFEUS
spectra, we show in Figure~1 a series of $20+37$ kK blackbody models
superposed on the data. While the absolute normalization (effective area)
of these fits differs somewhat from that of G\"ansicke et~al.\ because of
the different assumed white dwarf radius, distance, and the use of
blackbodies rather than white dwarf model atmosphere spectra, the trend
is the same, with maximum FUV continuum flux, and hence maximum project
spot area, occurring at $\phi
\approx 0.6$.

While the spot model of the continuum flux variations of AM~Her is simple
and appealing, there is an inconsistency with observations in that there
is no evidence in any of the ORFEUS spectra of the Lyman absorption lines
expected in the spectrum of a high-temperature, high-gravity stellar
atmosphere. This problem of missing photospheric absorption features is
not restricted to the FUV: the Lyman $\alpha $ absorption line is weak in
{\it HST\/} GHRS spectra (\cite{gan98}), the O~VI $2s$, $2p$ absorption
edges expected in the photosphere of the soft X-ray emission region are
not detected in {\it EUVE\/} SW spectra (\cite{pae96}), and the Ne~VI
$2s^22p$ and $2s2p^2$ absorption edges tentatively identified in {\it
EUVE\/} SW spectra acquired in 1993 September (\cite{mau95}; \cite{pae96})
are subtle at best in higher signal-to-noise ratio spectra acquired in
1995 March (\cite{mau98}). The standard explanation for the lack of
absorption features---a shallow photospheric temperature gradient caused
by irradiation by the hot, post-shock gas (\cite{tee94})---is not
entirely satisfactory for two reasons. First, the run of effective
temperature with optical depth must be fine-tuned so that emission lines
are not produced. Second, ``deep heating'' of the photosphere by blobby
accretion diminishes the fraction of the accretion luminosity in hard
X-rays and hence decreases the importance of irradiation. Detailed work
on the irradiation model is required to determine the severity of these
concerns.

\subsection{Emission Lines}

Accompanying the continuum flux variations are variations in the flux
and radial velocity of the emission lines. In what follows, we
concentrate on the emission lines longward of $\lambda = 1000$~\AA\ where
the spectra and hence the line fluxes and positions are not adversely
affected by the high background and consequent low signal-to-noise ratio.
To determine the parameters of the O~VI $\lambda\lambda 1032$, 1038
doublet, we fitted the spectra in the neighborhood of the line ($\lambda
= 1035\pm 10$~\AA ) with a linear continuum plus four Gaussians
corresponding to the broad and narrow components of each member of the
doublet, with the separations of the doublets constrained to the known
wavelength difference, and the widths of each component constrained to
be the same (10 free parameters). The C~III $\lambda 1176$ and He~II
$\lambda 1085$ lines could be and hence were fitted with a linear
continuum plus a single Gaussian (5 free parameters).

Figure~3 shows the fits of the O~VI lines, with both the data and the
model binned to a resolution of 0.1~\AA \ and smoothed with a 5-point
triangular filter. The figure demonstrates that a Gaussian
parameterization of the O~VI emission line components is adequate for
most of the spectra, and that, thanks to the high spectral resolution of
the Berkeley spectrograph, reasonably unique decompositions can be made
of these complex line profiles. The fitting parameters for the broad and
narrow components of the O~VI lines have been converted into physical
quantities (flux, velocity, FWHM) and are listed in Table~2. The
velocities of the broad and narrow components of the line are shown in
Figure~4 and have been fitted to a sinusoidal function $v=\gamma + K\,
\sin 2\pi (\phi-\phi_0)$ with the parameters shown in Table~3. Figure~4
demonstrates that the fit to the radial velocities of the narrow
component of the line is not good ($\rm\chi^2/dof=89/3$), possibly
because of residual variations in the zero point of the wavelength scale
of the spectrograph, but that at $K\approx 60~\rm km~s^{-1}$, its radial
velocity amplitude is on the low side of, but consistent with, the range 
of $K=60$--$150~\rm km~s^{-1}$ for the narrow IR and optical emission
lines measured on previous occasions (\cite{gre77}; \cite{you79};
\cite{cro81}). On the other hand, the radial velocity amplitude of the
broad component of the line is $K\approx 400~\rm km~s^{-1}$, which is
high compared to the range of $K=150$--$310~\rm km~s^{-1}$ for the broad
IR and optical emission lines. The similar amplitude and phase of the
radial velocity variations of the C~III and broad O~VI emission lines
argue that these lines originate in the same stream of gas within the
binary. In contrast, the He~II radial velocity is quite different:
compared to the C~III and broad O~VI lines, the He~II line has a similar
phase offset, but a large $\gamma $ velocity and half the radial velocity
amplitude. The He~II radial velocity behaves as if the stream of gas
responsible for this emission line is hidden from view when it is
directed toward the observer.

\section{Interpretation}

To understand the above results, it is necessary to establish the
phasing of the secondary and white dwarf and the relative orientation
of the X-ray emission region (these and other aspects of the binary are
summarized in Fig.~5). The phasing of the secondary is determined
relative to the magnetic ephemeris by the radial velocity of the star's
IR absorption lines. Leaving aside uncertain corrections for the
asymmetric irradiation of the inner face of the secondary by the X-ray
emission region, inferior conjunction of the secondary (blue to red
zero crossing of the absorption lines) occurs at $\phi\approx 0.65$
(\cite{you79}; \cite{sou95}). The geometry of the X-ray emission region
is determined by quasi-simultaneous {\it EUVE\/}, {\it ROSAT\/}, and
{\it ASCA\/} observations of AM~Her in 1995 March (\cite{mau98}, see also
\cite{pae96} and \cite{isi97}). At hard X-ray energies ($E\approx 4$--10
keV) where the hard ``thermal brems'' component dominates the X-ray
spectrum, the X-ray light curves are single peaked, with a broad, flat
maximum centered at $\phi\approx 0.6$ and an eclipse (of the X-ray
emission region by the body of the white dwarf) centered at $\phi
\approx 0.11$. At soft X-ray energies ($E\lax 1$~keV) where the soft
``blackbody'' component dominates the X-ray spectrum, the light curves
are more complex, with a ``bright'' phase from $\phi\approx 0.3$--0.6, a
``dim'' phase from $\phi\approx 0.6$--1.0, and the same eclipse observed
at higher energies. The {\it ROSAT\/} HRI ($E\approx 0.1$--2.0 keV) light
curve tracks the {\it EUVE\/} DS and SW ($E\approx 0.1$--0.2 keV) light
curves over most binary phases, but from $\phi\approx 0.3$--0.6, excess
absorption is indicated by a deficit of EUV flux.

Given that the FUV and hard X-ray light curves both peak at $\phi\approx
0.6$, it is apparent that the $\approx 37$~kK hot spot we invoked in \S 3
to describe the FUV continuum flux variations is associated with the X-ray
emission region. The size of these emission regions depends sensitively 
on their effective temperatures, but for what seem to be reasonable
assumptions, the FUV emission region is significantly larger than the
X-ray emission region. Specifically, blackbody fits to {\it EUVE\/}
spectra yield $T=240$--290 kK and $f_{\rm X}=2$--$5\times 10^{-3}$ for
the soft X-ray emission region (\cite{pae96}), while fits to the {\it
IUE\/} and ORFEUS spectra yield $T\approx 37$ kK and $f_{\rm FUV}\sim
0.1$ for the FUV emission region (\S 3 and \cite{gan95}). A larger FUV
emission region is also consistent with the fact that the FUV continuum
flux is never fully eclipsed by the body of the white dwarf, whereas
the X-ray flux is. These emission regions are offset relative to the
secondary by $\Delta\phi\approx 0.05$, consistent with the azimuth of
the magnetic pole of the white dwarf ($\psi=0.046\pm 0.033$; \cite{cro88},
but we have corrected a mistake therein [$\psi=\phi_{\rm s}-\phi_{\rm
c}=0.653\pm 0.029 - 0.607\pm 0.016 = 0.046\pm 0.033$, not $0.087\pm
0.030$]).

Since the maximum redshift of the broad O~VI, C~III, and He~II emission
lines occur at approximately the same phase as the maximum visibility
of the FUV and X-ray emission regions, it is reasonable to conclude
that these lines are produced in the funnel of gas directed towards and
falling onto the FUV and X-ray bright accretion region in the vicinity of
the magnetic pole of the white dwarf. Half a binary orbital period later,
the X-ray emission region is eclipsed by the body of the white dwarf,
and the broad O~VI, C~III, and He~II emission lines attain their maximum
blueshifts. The pathological behavior of the radial velocities of the
He~II emission line ($\sim 0~\rm km~s^{-1}$ maximum blueshift) suggests
that the portion of the accretion funnel responsible for this emission
line is eclipsed along with the X-ray emission region by the body of
the white dwarf. The excess absorption from $\phi\approx 0.3$--0.6 is
presumably due to the accretion stream and funnel, which lies along the
line of sight at these binary phases. On the other hand, the cause of
the ``bright'' and ``dim'' phases of the EUV and soft X-ray light curves
is a mystery, deepened by the fact that in 1991 April (\cite{gan95})
the ``dim'' phase of the soft X-ray light curve {\it preceded\/} the
``bright'' phase, the reverse of what was observed in 1993 September
(\cite{pae96}) and 1995 March (\cite{mau98}). The solution of this puzzle
is left as an exercise to the reader.
 
The physical conditions of the FUV line-emitting gas are constrained by
a number of pieces of information. In gas illuminated by a mixture of
hard and soft X-rays, O~VI exists over a range of ionization parameters
$\xi\approx 30$--60 and temperatures $T\approx 20$--100~kK, peaking at
$\xi\approx 35$ and $T\approx 20$~kK (\cite{kal82}, model 5; pure soft
X-ray illumination gives a similar result according to the code described
by \cite{ray93}). Support for the inferred temperature range is provided
by the mean C~III 977/1176 line ratio of $\sim 2$, which requires $T\sim
50$~kK (\cite{kee92}; \cite{kee97}). Over the inferred range of ionization
parameters, helium makes the transition from He~II to He~III, with He~III
dominant above $\xi\approx 35$. Similarly, the dominant ionization stage
of carbon should be C~IV--VI, not C~III, but this is consistent with
the fact that C~IV $\lambda 1550$ in the UV is much brighter than C~III
$\lambda 977$, $\lambda 1176$ in the FUV (\cite{ray79}; \cite{hei88};
\cite{gre96}). The density of the O~VI narrow-line region is $n_{\rm nlr}
=L/\xi r_{\rm nlr}^2\sim 3\times 10^{10}\, L_{33.5}~\rm cm^{-3}$, where
$r_{\rm nlr} =5.3\times 10^{10}$ cm is the distance from the white dwarf
to the face of the secondary and $L_{33.5}$ is the X-ray luminosity in
units of $3\times 10^{33}~\rm erg~s^{-1}$ (\cite{pae96}). The density of
the O~VI broad-line region is uncertain because its distance from the
white dwarf is uncertain, but at $r_{\rm blr}\gax 1\times 10^{10}$~cm
(where the free-fall velocity is $\lax 1300~\rm km~s^{-1}$, much higher
than the O~VI radial velocity amplitude), $n_{\rm blr}\lax 1\times
10^{12}\, L_{33.5}~\rm cm^{-3}.$

Surprisingly, even though the O~VI broad-line region is a factor of
$\sim 30$ times denser than the O~VI narrow-line region, the ratio $R$
of the O~VI line intensities shown in Table~2 demonstrates that it is 
the narrow-line region ($R\approx 1$), {\it not\/} the broad-line region
($R\approx 2$), which is optically thick. If the broad-line region were
fully optically thin, it would be equally bright when viewed from any
orientation (i.e., any orbital phase); if optically thick and elongated,
it would be bright twice per orbital period and dim twice per orbital
period; in fact, the O~VI broad-line flux varies sinusoidally, with a
maximum flux at $\phi\approx 0.2$ and a minimum at $\phi\approx 0.7$
(Fig.~6). The cause of this variation is unknown. Puzzling too is the
variation with phase of the O~VI narrow-line flux. From our association
of the O~VI narrow-line region with the irradiated face of the secondary,
one would expect minimum O~VI narrow-line flux at inferior conjunction
of the secondary at $\phi\approx 0.65$ and a broad maximum $180^\circ $
later, but this is not the case.

We close by considering the relationship between the broad Balmer and
O~VI line-emitting gas. Stockman et~al.\ (1977) found that the Balmer
lines arise in an optically thick region with a density  of $10^{13} <
n({\rm cm^{-3}}) < 2\times 10^{14}$, which is factor of $\gax 10$--200
times higher than the density of the O~VI broad-line gas. That the
Balmer and O~VI emission line regions are at least roughly cospatial is
indicated by the similar amplitude and phasing of their radial velocity
variations (Table~3, \cite{pri77}, and \cite{cow77}). The ratio of the
radial velocity amplitudes is $K_{\rm O~VI}/ K_{\rm Balmer}=1.6\pm 0.2$
and $\sim 1.1$ for the data of Priedhorsky (1977) and Cowley \& Crampton
(1977), respectively. Assuming that the gas is in free-fall onto the
white dwarf, $K_{\rm O~VI}/K_{\rm Balmer}<2$ implies $r_{\rm Balmer}/
r_{\rm O~VI}<4$. By mass conservation, $n_{\rm Balmer}/n_{\rm O~VI}<2$
if the stream cross section is constant with radius, and $>1/32$ if the
stream cross section scales like the dipolar field as $r^3$. The
ionization parameter of the Balmer emission-line region is then $\xi\gax
1$, which does not allow H~I to dominate over H~II. Consequently, the
accretion flow must be clumped for the O~VI and Balmer emission lines to
coexist, consistent with the idea of clumpy accretion as an explanation
for the soft X-ray excess (\cite{kui82}) and for the variability observed
in hard X-rays and at optical wavelengths (\cite{bea97}).

\acknowledgments

We thank all those involved with making the ORFEUS-SPAS II mission a
success: the members of the satellite and instrument teams at Institute
for Astronomy and Astrophysics, University of T\"ubingen; Space Science
Laboratory, University of California, Berkeley; and Landessternwarte
Heidelberg-K\"onigstuhl; the flight operations team; and the crew of
STS-80. Special thanks are due to Van Dixon for his heroic efforts
scheduling the ORFEUS guest investigator program and particularly for
securing the excellent phase coverage of AM~Her. F.\ Keenan is warmly
thanked for supplying us with C~III level population and line intensity
data. The anonymous referee is acknowledged for helpful comments. C.\ W.\
M.'s contribution to this work was performed under the auspices of the
U.S.~Department of Energy by Lawrence Livermore National Laboratory under
contract No.~W-7405-Eng-48.

\clearpage % force page break

%redefine ! to use in the tables as a spacer: 

\newdimen\digitwidth
\setbox0=\hbox{\rm0}
\digitwidth=\wd0
\catcode`!=\active
\def!{\kern\digitwidth}

% Table 1
%---------------------------------------------------------

\begin{table*}
\begin{center}
\begin{tabular}{ccc}
\multicolumn{3}{c}{TABLE 1} \\
\multicolumn{3}{c}{Journal of Observations} \\
\tableline \tableline
Start Date& Exposure& \\
($\rm HJD-2450000$)& (s)& $\phi$\\
\tableline
413.67401&  1876&  !0.332--!0.501\\
413.73574&  1951&  !0.811--!0.986\\
414.68375&  1953&  !8.164--!8.339\\
414.74966&  1772&  !8.675--!8.834\\
415.82732&  1990&  17.034--17.213\\
415.89073&  2007&  17.526--17.706\\
\tableline
\end{tabular}
\end{center}
\end{table*}

\clearpage % force page break

% Table 2
%---------------------------------------------------------

% Table 2 must be \LaTeX ed separately, printed sideways, and inserted
% here; increment page counter so that this can be accomplished.......

\addtocounter{page}{1}

% Table 3
%---------------------------------------------------------

\begin{table*}
\begin{center}
\begin{tabular}{lcccc}
\multicolumn{5}{c}{TABLE 3} \\
\multicolumn{5}{c}{Radial Velocities\tablenotemark{a}} \\
\tableline \tableline
    &          & $\gamma $&         $K$&                        \\
Line& Component& $\rm (km~s^{-1})$& $\rm (km~s^{-1})$& $\phi _0$\\
\tableline
O VI  $\lambda\lambda 1032$, 1038& N!& $!-22\pm !6$&  $!57\pm !9$&  $0.700\pm 0.022$\\
O VI  $\lambda\lambda 1032$, 1038& B!& $!-49\pm 14$&  $412\pm 19$&  $0.297\pm 0.007$\\
He II $\lambda        1085$      & B?& $+176\pm 18$&  $180\pm 29$&  $0.391\pm 0.018$\\
C III $\lambda        1176$      & B!& $!!+6\pm 19$&  $480\pm 29$&  $0.294\pm 0.008$\\
\tableline
\end{tabular}
\tablenotetext{a}{$v=\gamma + K\, \sin\, 2\pi (\phi - \phi _0)$.}
\end{center}
\end{table*}

\clearpage % force page break

% References
%---------------------------------------------------------

\clearpage % force page break

% Figure captions
%---------------------------------------------------------

\figcaption%[figure1.ps]
{ORFEUS spectra of AM Her ordered by binary phase. Each successive 
spectrum is offset by 1.5 flux density units. Two-component ($20+37$ kK)
blackbody models are shown by the light-colored smooth curves.
\label{fig1}}

\figcaption%[figure2.ps]
{Mean flux density at $\lambda = 1010\pm 5$~\AA \ in units of
$\rm 10^{-12}~erg~cm^{-2}~s^{-1}~\AA ^{-1}$ as a function of binary
phase.
\label{fig2}}

\figcaption%[figure3.ps]
{Regions of the spectra containing the O~VI doublet showing Gaussian
fits to the broad and narrow components. Panels are ordered by relative
binary phase.
\label{fig3}}

\figcaption%[figure4.ps]
{Radial velocity of the broad ({\it boxes\/}) and narrow ({\it 
crosses\/}) components of the O~VI doublet as a function of binary
phase.
\label{fig4}}

\figcaption%[figure5.ps]
{Schematic diagram of AM~Her showing the velocity vectors and inferred
locations of the FUV broad- and narrow-line regions and the phases of 
the X-ray eclipse, EUV flux deficit, maximum FUV and hard X-ray continuum
flux, maximum O~VI broad-line flux, and maximum blueshift of the He~II
and broad O~VI lines.
\label{fig5}}

\figcaption%[figure6.ps]
{Flux in units of $\rm 10^{-12}~erg~cm^{-2}~s^{-1}$ of the broad ({\it
boxes\/}) and narrow ({\it  crosses\/}) components of the O~VI doublet
as a function of binary phase.
\label{fig6}}

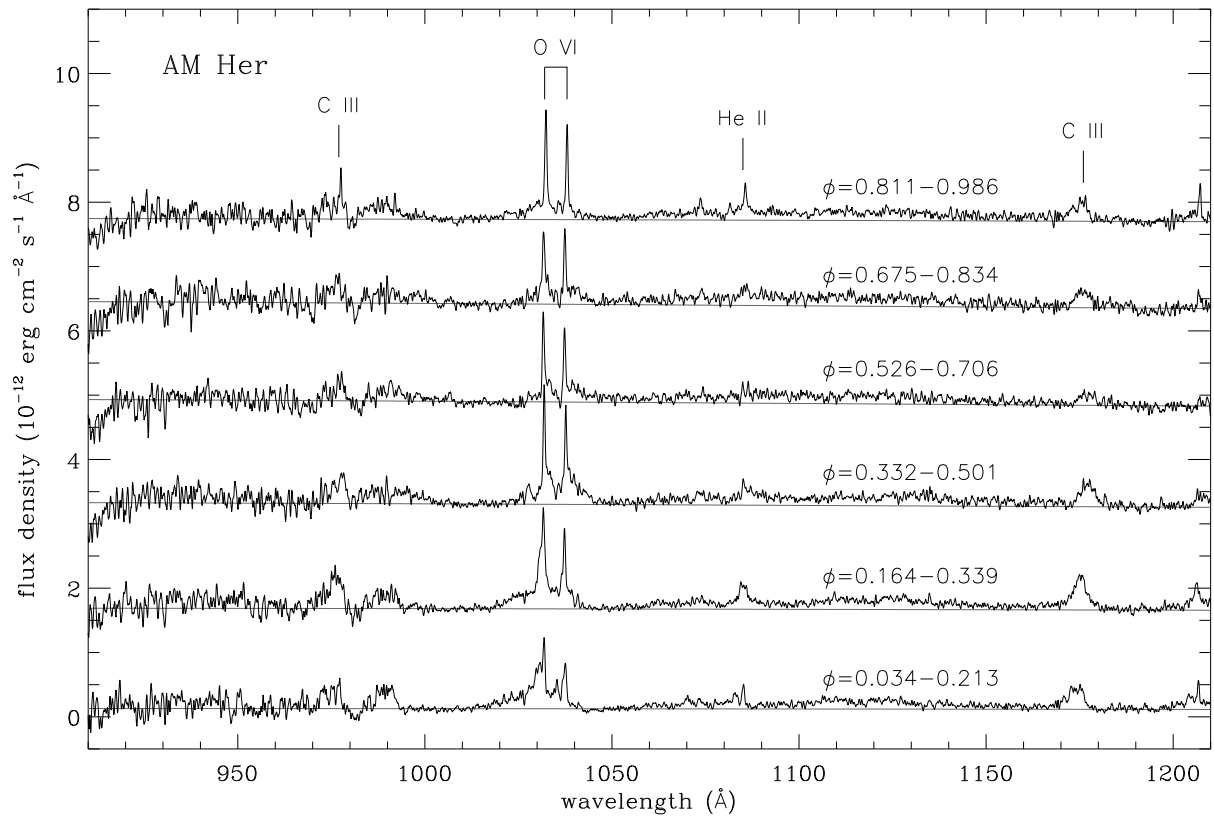
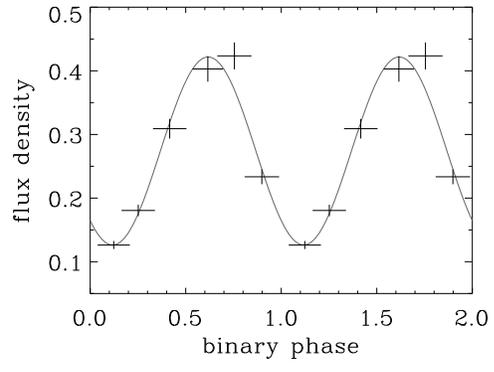
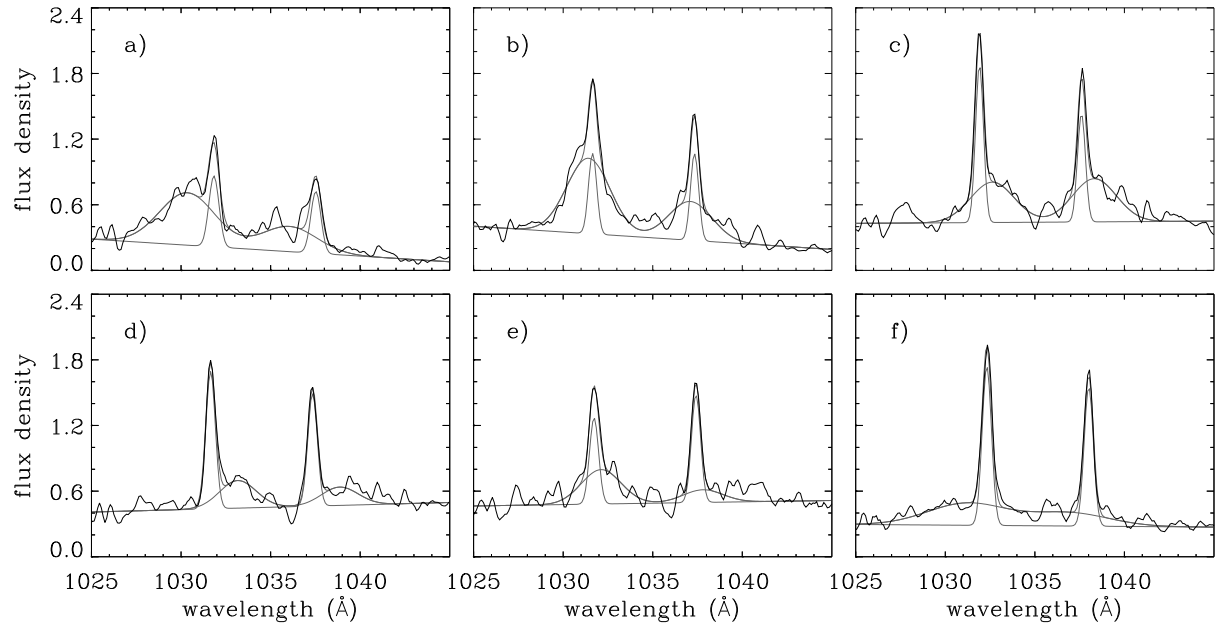
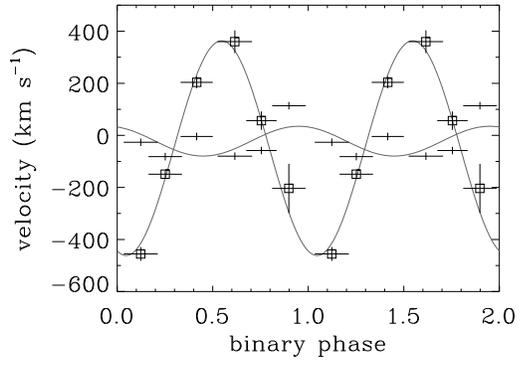
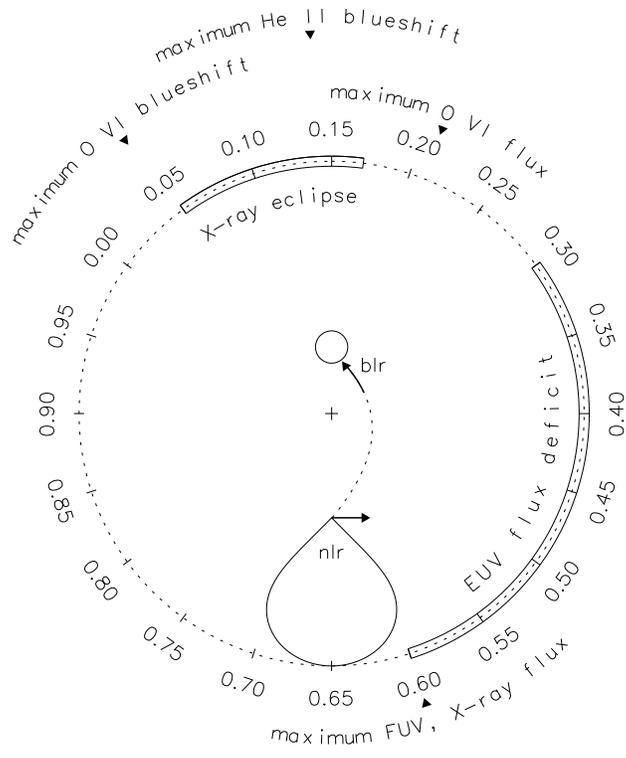
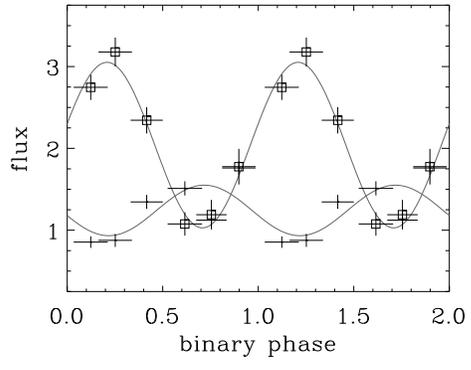
\end{document}